\documentclass[12pt]{article}
\usepackage{amsmath,amsfonts,amssymb,a4,epsfig}

\newcommand{\R}{{\mathbb{R}}}

\newcommand{\Z}{{\mathbb{Z}}}

\def\ha{\frac{1}{2}}

\def\pa{\partial}
\def\ra{\rightarrow}
\def\preuve{\begin{proof}} 

\def\gb{\beta}

\def\gd{\delta}

\def\gl{\lambda}
\def\go{\omega}

\def\san{San V{\~u} Ng{\d o}c}
\def\OPD{~$\Psi DO$}

\newtheorem{defi}{Definition}[section]

\newtheorem{lemm}{Lemma}[section]

\newtheorem{rem}{Remark}[section]
\newtheorem{coro}{Corollary}[section]
\newtheorem{theo}{Theorem}[section]

\newenvironment{demo}{\noindent {\it Proof.--}
      \begin{quotation}\noindent}{\end{quotation}\hfill$\square $}

\begin{document}

\title{A semi-classical inverse problem I:\\
Taylor expansions.\\
{\normalsize (to Hans Duistermaat for his 65 birthday)}}
\author{Yves Colin de Verdi\`ere\footnote{Institut Fourier,
 Unit{\'e} mixte
 de recherche CNRS-UJF 5582,
 BP 74, 38402-Saint Martin d'H\`eres Cedex (France);
yves.colin-de-verdiere@ujf-grenoble.fr}
~ \& Victor Guillemin\footnote{Math. Dep. MIT; vwg@mit.edu}}

%\date{}

\maketitle

\begin{abstract}
In  dimension 1, we show that the
 Taylor expansion of a ``generic'' potential near a non degenerate
critical point can be recovered from the knowledge of the
semi-classical
spectrum of the associated
Schr\"odinger operator
near the corresponding critical value. Contrary to the work
of previous authors, {\bf we do not assume that the potential is even.}
The classical Birkhoff normal form does not contain
enough information to determine the potential,
 but the quantum Birkhoff normal form
does\footnote{
This work  started from  discussions we had  during the Hans conference
in Utrecht  (August
2007). The proofs were completed independently by both authors
two  months later. We 
decided then  to write a joint  paper.}.

In a companion  paper \cite{CV3}, the first author shows how the
potential
itself is, without any analyticity assumption  and under some mild
genericity
hypotheses, determined by the semi-classical spectrum. 

\end{abstract}
\section{Introduction}
In this paper\footnote{
Many thanks to Fr\'ed\'eric Faure for discussions
 and his computations}, we  will only  consider a   configuration space 
of dimension $1$.

Let us consider a (classical) Hamiltonian
\[ H(x,\xi)=\ha\xi^2 +V(x) ~\]
with $V(0)=E_0,~V'(0)=0,~V''(0)=\pm 1$(\footnote{Assuming  that
$V''(0)=\pm 1$ does not affect the
results below, because $a_2=V''(0)/2$ is known from the first
eigenvalue if $a_2>0$ and from the density of states if 
$a_2<0$}). We have 
\[ H(x,\xi)\equiv E_0 +  \Omega_\pm  +\sum_{j=3}^\infty  a_j x^j \]
with $\Omega_\pm  =\ha (\xi^2 \pm x^2 )$.
The Hamiltonian $H$ can be quantized as a Schr\"odinger operator
$\hat{H}=-\ha \hbar^2\frac{d^2}{dx^2}+V(x)$
where the Taylor expansion of $V$ at $x=0$ is
$E_0 + \sum_{j=2}^\infty  a_j x^j $ with $a_2=\pm \ha $.
This operator  admits a semi-classical Birkhoff normal form 
\cite{San-C}
(denoted the QBNF)
at the origin of which the Weyl symbol is a formal power series 
of the form
 \begin{equation} \label{equ:qbnf}
 B\equiv \Omega _\pm + \sum_{2j+k\geq 2 }
 b_{j,k}\hbar ^{2j}\Omega ^k_\pm~. \end{equation}
In this paper, we are  interested in the following ``inverse
spectral problem'':

{\bf does the QBNF,  given in (\ref{equ:qbnf}),
of the Schr\"odinger operator determine  
the Taylor series of $V$?}

We cannot hope for a positive answer , because $V(x)$
and $V(-x)$ give the same QBNF. Moreover

\begin{rem}{\bf the classical BNF does not suffice to determine
 the 
Taylor expansion of  V at x=0}

Let $y=f(x)=x+O(x^2)$ be an analytic function
whose local inverse near $0$ is of the form
$x=y+g(y)$ with $g$ an {\rm even} function.
Then the Hamiltonian 
$ H_f=\ha \left( \xi^2 + f(x)^2 \right) $ is classically
conjugate to $\Omega _+$ near the origin, in particular all its
trajectories
are of period $2\pi$:
it is enough to show that the action integrals
$I(E)=\int _{\xi ^2 + f(x)^2 \leq 2E } dx d\xi $ are the same;
using the change of variable $x=y+g(y)$, we get
$I(E)=\int_{\xi ^2 + y^2 \leq 2E }  (1+g'(y))dy d\xi $
and using the fact that $g'$ is {\rm odd} we get the result.
A simple example is
$V(x)=\ha \left( \sqrt{1+2x}-1 \right) ^2 $.
This result is reminiscent of  the well known result for Zoll surfaces
in Riemannian geometry \cite{besse}.

However, an even potential can be determined by the classical BNF,
as  a consequence of a result of N. Abel
\cite{abel}\footnote{We are grateful to Hans for pointing this out to us}.
\end{rem}

Our main result is:
\begin{theo}
The coefficients $\pm a_3 $ and $a_4$ are determined from
$b_{0,2}$ and $b_{1,0}$
by the formulas:
\[a_3=\pm \sqrt{b_{1,0}},~a_4=\frac{2}{3}b_{0,2}+\frac{5}{2} b_{1,0}~.
\] 

If {\bf  $a_3$ does not vanish,}
all $a_j$'s are determined from the $b_{0,k}$'s and the $b_{1,k}$'s
once we  have chosen the sign of $a_3$.
\end{theo}
This result is reminiscent of the much more sophisticated results
by Zelditch on the Kac problem \cite{Ze}.
If we use a (trivial) particular case of the result of \cite{G-P-U}, we get
\begin{coro}
If we know the asymptotic expansions 
of the eigenvalues $\gl_n(\hbar )$
for all $n's$ of a Schr\"odinger operator near the
minimum $x=0$ of the potential and $V''(0)>0$,
we know the value of $V'''(0)$ and, if
$V'''(0)\ne 0$, the  Taylor expansion of the potential at that point.
\end{coro}
In fact, we have the more precise result:
\begin{coro}
From the knowledge of the $N$ first eigenvalues
of $\hat{H}$ modulo $O(\hbar^{2N})$,
one can recover the Taylor expansion of $V$ to order
$2N$.
\end{coro}

A similar result holds for a local non degenerate maximum of $V$
using the ``density of states'' techniques. This is the content of Section
\ref{sec:density}:
\begin{coro}
If $E_0$ is a non degenerate local maximum of $V$ and $0$ is the only
critical
point of $V$ on the set $V=E_0$,
the knowledge of the semi-classical spectrum of
$\hat{H}$ in some intervall $]E_0, E_1 [$ (or
 $]E_1, E_0 [$)  determines $V'''(0)$
and,  provided that $V'''(0)\ne 0 $, the Taylor expansion of $V$
at $x=0$.
\end{coro}
and also in the case of a local minimum (Section \ref{sec:minimum}):
\begin{coro}
If $E_0$ is a non degenerate local minimum of $V$ and $0$ is the only
critical
point of $V$ on the set $V=E_0$,
the knowledge of the semi-classical spectrum of
$\hat{H}$ in some   interval $]E_1, E_2 [$, 
with $E_1<E_0<E_2$,
  determines $V'''(0)$
and,  provided that $V'''(0)\ne 0 $, the Taylor expansion of $V$
at $x=0$.
\end{coro}

Knowing the semi-classical spectrum as a function
of $\hbar $ seems  to be an huge amount of information. 
As was showed  in \cite{CV1}, this is however 
the case for the effective Hamiltonians driving the propagation
of waves inside a stratified medium.

\section{A counterexample for a general Hamiltonian}

The QBNF of a general Hamiltonian, {\bf independent of
$\hbar $},  $H(x,\xi)=\Omega _\pm + O(3)$
is not enough to know the Taylor expansion of $H$ at the singular point.
It is enough to consider
$H=\ha \left( (\xi-3x^2)^2 + x^2\right) $
which is gauge equivalent to
$\Omega _+$ by the gauge transform $u\ra ue^{ix^3}$.
\section{Review of the Moyal product}
The Moyal product is the product rule of symbols of 
Weyl quantized \OPD's, it is given by:
\[ a\star b \equiv\sum _{j=0}^\infty
\frac{1}{j!}\left(\frac{\hbar}{2i}\right)^j \{ a, b\}_j \]
with
%with $P={\buildrel \leftarrow \over \nabla} _{ \xi}
%{\buildrel \rightarrow \over \nabla} _{ x}-
%{\buildrel \leftarrow \over \nabla} _{x}
%{\buildrel \rightarrow \over\nabla} _{ \xi  } $.
\[ \{ a, b\}_j:=\sum _{p=0}^{j}\left( \begin{array}{c} p\\ j
  \end{array}
\right)(-1)^p \pa _x^p \pa _\xi ^{j-p} a \pa _x^{j-p}\pa _\xi ^p b ~.   \]
We will also use the {\it Moyal bracket},
\[ [ a,b]^\star =a\star b-b\star a ~.\]
We have
\[ \frac{i}{\hbar} [ a,b]^\star \equiv 
\sum _{j=0}^\infty \frac{1}{2j +1 !} \left(\frac{\hbar}{2i}\right)^{2j }
\{a,  b \}_{2j+1}~.\]
 In particular, 
$\{ a,b \}_1=a_\xi b_x -a_xb_\xi $ is the Poisson bracket and 
\[ \{ a,b \}_3=a_{\xi \xi \xi}b_{xxx}-3 a_{\xi \xi x}b_{xx \xi}+
3 a_{\xi x x}b_{x \xi \xi } -a_{xxx}b_{\xi \xi \xi}~.\]
We have:
\[ \frac{i}{\hbar} [ a,b]^\star \equiv \{ a, b\}_1 -\frac{\hbar ^2}{24}
\{ a, b\}_3 +\frac{\hbar ^4}{1920} \{ a, b\}_5+\cdots ~.\]

\section{The Weyl algebra }
The ``Weyl algebra'' which consists
of formal power series in $\hbar $ and $(x,\xi) $
\[ W=\sum_{j=2}^\infty W_j \] 
where $W_j $ is the space of polynomials
in $(x,\xi) $ and $\hbar $ of total degree $j$
and  the degree of $x^l\xi^m \hbar^{n}$
is $l+m+2n$.
$W$ is a graded algebra for the Moyal product:
we have $W_j\star W_k \subset W_{j+k} $ and   hence  
  $\frac{i}{\hbar }[W_j,W_k]^\star \subset W_{j+k-2}$.
Moreover, if we define $W_j^+$ as the subspace of $W_j$
which is generated by monomials of even degree in $\hbar$, we
have
\[ \frac{i}{\hbar }[W_j^+,W_k^+]^\star \subset W_{j+k-2}^+ ~;\]
we will define $W^+=\sum _{j=3}^\infty W_j^+ $ which is a Lie
algebra for the bracket  $\frac{\hbar }{i}[.,.]^\star$. $W^+ $
is the (formal) Lie algebra of FIO's which are tangent to the identity  at the
 the origin.
The grading is obtained by looking at the action on the
(graded) vector space of {\it symplectic spinors}:
if $F\equiv \sum_{j=0}^\infty \hbar ^j F_j (X) $ with $F\in {\cal
  S}(\R)$,
we define
$f_\hbar (x)=\hbar^{-\ha}F(x/\hbar)  $ whose microsupport
is the origin. $W^+ $ acts on this space of functions
in a graded way as differential operators of infinite degree:
if $w\in W$, $w.f={\rm OP}_\hbar (w)(f) $.

\section{Moyal versus functional QBNF}

There are two different  QBNF:
\begin{itemize}
\item The first one  is a {\it Weyl symbol}
$B\equiv \sum b_{j,k}\hbar^{2j}\Omega ^k$
as before,
\item  The second  one is an {\it operator} which is a formal power series
of the harmonic oscillator $\hat{\Omega }$ of 
the form
$\hat{B} \equiv \sum \widehat{b}_{j,k}\hbar^{2j}\hat{\Omega} ^k$.
\end{itemize}
The second one is the Weyl quantization of the first. So they 
are equivalent. The equivalence can be made  explicit in both 
direction by computing
${\rm Op}_{\rm Weyl} (\Omega ^k )$ or 
the Weyl symbol of $\hat{\Omega}^k $. 
The functional form is useful in order to compute successive
approximations
of the eigenvalues, while the Weyl form is easier to compute using
the Moyal product. 
\section{Useful Lemmas}
The following result is classical: 
\begin{lemm} \label{lemm:homol}
The equation 
$\{ \Omega _\pm , P\}_1 =Q $
where $Q$ is a given  homogeneous polynomial of degree $N$ 
has a solution $P$, a   homogeneous polynomial of degree $N$,
\begin{itemize}
\item if $N$ is odd
\item if $N=2N'$ is even and
$c_\pm {(Q)}=0$ where  $c_\pm $ is a linear form on
the space of  homogeneous polynomials of degree $N$ which satisfies 
$c_\pm (\Omega _\pm ^{N'})=1$.
In particular, given $Q$, 
 the equation $  \{ \Omega _\pm , P\}_1 =Q-c_\pm (Q) \Omega _\pm ^{N'} $
has a solution.
\end{itemize}
\end{lemm}
\begin{rem} In the case $\Omega _+$, $c_+(Q)\Omega _+^{N '}$ is the
  average of $Q$ under the natural  action of $S^1$ on homogeneous
  polynomials
of degree $2N'$.
\end{rem}
\begin{defi}\label{def:sigma}
We will denote by $\Sigma _{2N-1}^\pm$ the homogeneous polynomial
of degree $2N-1$ which satisfies
\[ \{\Omega _\pm, \Sigma _{2N-1}^\pm    \}_1= x^{2N-1}~.\]
\end{defi}
\begin{lemm}\label{lemm:Sigma}
We have 
\[ \Sigma _{2N-1}^\pm=-\left(\pm  x^{2N-2}\xi +\frac{2N-2}{3}x^{2N-4}\xi^3
+\cdots \right)  ~.\] 
\end{lemm}
We can also check the:
\begin{lemm} \label{lemm:x2N}
 The polynomials $x^{2N'}$ are not  Poisson brackets of
the form 
$x^{2N'}=\{ \Omega _\pm , P \}_1 $, i.e. $c_\pm (x^{2N'})\ne 0$.
\end{lemm}  

\section{The QBNF}
In order to reduce to the QBNF, we will 
use automorphisms of $W^+$ of the form
\[ H\ra H_S={\rm exp}(iS/\hbar)\star H \star {\rm exp}(-iS/\hbar)=
{\rm exp}\left( \frac{i}{\hbar }ad (S)^\star \right)  H\] 
with $S=S_3 +S_4 +\cdots \in W^+$.
 We get:
\[ H_S =H+\frac{i}{\hbar} [S,H]^\star
+\cdots + \frac{1}{k!}
\left( \frac{i}{\hbar}\right)^k\overbrace{ [S,[S,\cdots,[}^{k~{\rm brackets}}
S,H]^\star]^\star\cdots ]^\star
+\cdots ~,\]
which is a convergent formal power series whose
$k-$th term is of degree $\geq k+2$.  
The brackets will be calculated using the Moyal bracket.
We  remark that the terms of degree $0$ in $\hbar $ give 
the calculation of the classical BNF (denoted CBNF)
where the brackets are now just Poisson brackets. 

\section{The first terms}
Let us consider 
$V(x)=\ha x^2 + ax^3 +bx^4 + \cdots $
whose QBNF is
$\Omega + A\Omega ^2 +B\hbar ^2 + O(6) $
where $O(6)$ means terms of degree $\geq 6$ in the Weyl algebra.
Here we assume $\Omega =\ha (\xi^2 +x^2)$.
Our first result is:
\begin{theo}
\[ A=-\frac{15}{4} a^2 +\frac{3}{2}b,~B=a^2 ~.\]
\end{theo}
{\it The calculation:} we  start with
$S=S_3 + S_4 $ where $S_3(x,\xi)$(resp. $S_4(x,\xi)$)
is a homogeneous polynomial of degree $3$(resp. $4$).
There is  no need to put terms in $\hbar ^2$ in $S_4$
because they would be of the form $c\hbar^2$ which is in the
center. 
We have then:
\[ {\rm exp}(\frac{i}{\hbar}[S,.]^\star)H=
\Omega + \frac{i}{\hbar}[S,H]^\star+\ha \left(\frac{i}{\hbar}\right)^2
[S,[S,H]^\star]^\star+
0(6)~.\]
By identification of terms of degree $3$ and $4$ and  using
the expression of the Moyal bracket $[.,.]^\star$:
\[ \frac{i}{\hbar}[f,g]^\star =\{ f,g\}_1 -\frac{1}{24}\hbar ^2
\{f,g\}_3 + \cdots ~,\]
we get
the system of equations:
\[ \left\{ \begin{array}{crl}(3)& ax ^3+ \{ S_3, \Omega \}_1&=0\\
(4)& bx^4 + \{ S_3, ax^3 \}_1 +\{ S_4,\Omega \}_1
+\ha \{ S_3, \{ S_3, \Omega \}_1 \}_1 
-\frac{1}{24}\hbar^2 \{ S_3, ax^3 \}_3 &=A\Omega ^2 +B\hbar^2 
  \end{array} \right. \]
Using Equation (3), Equation (4) splits into 2 equations:
\[ \left\{ \begin{array}{crl}(4')&-\frac{1}{24} \{ S_3, ax^3 \}_3&=B  \\
(4'')& bx^4 +\ha  \{ S_3, ax^3 \}_1 +\{ S_4,\Omega \}_1
   &=A\Omega ^2 
\end{array} \right. \]
From Equation (3) and the formula for $\Sigma _{2N-1}$
given in Lemma \ref{def:sigma}, we get
\begin{equation} \label{equ:S3}
S_3= -a(x^2\xi + \frac{2}{3}\xi^3 )~.\end{equation}
From Equation (4'), we get $B=a^2$.
From Equation (4''), we get the value of $A$.

\section{The induction}

{\it We carry out  the proof in the case of $\Omega _+$
and $E_0=0$. The minus case is
similar.}

Let us start with
\[ H'=\Omega _+ + a_3 x^3 +\cdots +a_{2N-2}x^{2N-2}\] 
and $S'=S_3 +S_4 +\cdots +  S_{2N-2}$
with $S_j\in W_j$, 
so that 
\[ {\rm exp}\left( \frac{i}{\hbar} [S',.]^\star \right)
H'=\Omega _+ + B_4 +\cdots + B_{2N-2}+ R_{2N-1} +R_{2N}+ \cdots(:=B')
~,\]
with 
\begin{itemize}
\item $B_{2j} \in W_{2j}^+$  a polynomial in $\hbar^2  $
and $\Omega _+$
\item For $n=2N-1$ and $n=2N$,  $R_{n}\in W_n $.
\end{itemize}
In other words $S'$ generates the transformation which
converts $H'$ into its QBNF mod $O(2N-1)$. The polynomials
$H'$ and  $S'$ and the partial  QBNF $B'$
are known by the induction hypothesis.
We are now trying to get $S''=S_{2N-1}+ S_{2N}$ so that
$S=S'+S''$ converts $H=H'+ ax^{2N-1}+ bx^{2N}$ into  the QBNF
mod $O(2N+1)$. We will only consider the terms of degree $0$
and $2$ in $\hbar $. So we can split every polynomial $P_j$ in
$W_j^+$  into  $P_j=P_j^0 +\hbar ^2 P_j^2 + \cdots $
with $ P_j^2 $ of degree $j-4$ in $(x,\xi)$.

The equation to solve is:
\begin{equation}
\label{equ:ind} 
\begin{array}{rl}  {\rm exp}\left( \frac{i}{\hbar} [S'+S'',.]^\star \right)
(H'+ax^{2N-1}+ bx^{2N})=\\
  \Omega _+  + B_4 +\cdots + B_{2N-2}+
B_{2N}+& O(2N+1)~\end{array}\end{equation}
with $B_{2N}=b_{2N}^0 \Omega_+^N +b_{2N}^2 \hbar ^2 \Omega
_+^{N-2}+\cdots $.
We hope to recover $a$ and $b$ from $b_{2N}^0$
and $b_{2N}^2$ using what we know already at this step.
The left handside of Equation (\ref{equ:ind})  splits into:
\[  {\rm exp}\left( \frac{i}{\hbar} [S'+S'',.]^\star \right)
(ax^{2N-1}+ bx^{2N})=ax^{2N-1}+
bx^{2N}+\frac{i}{\hbar}[S_3,ax^{2N-1}]^\star +
O(2N+1)~,\]
and 
\begin{eqnarray*} \label{2N-1}
  {\rm exp}\left( \frac{i}{\hbar} [S'+S'',.]^\star \right)H'=
B'+ \frac{i}{\hbar}[S'',\Omega _++a_3x^3]^\star+ \\
+\ha \left( \frac{i}{\hbar} \right)^2  \left( [S_{2N-1},[S_3,\Omega
  _+]^\star]^\star 
+ [S_{3},[S_{2N-1},\Omega _+]^\star]^\star\right)
+0(2N+1)~.\nonumber \end{eqnarray*}
So that, we get
\begin{itemize}
\item
%-----------------------------------------------------------
%degree 2N-1
{\bf  In degree $2N-1$:}
\[ ax^{2N-1}+\{ S_{2N-1}^0,\Omega _+    \}_1 +R_{2N-1}^0=0  \]
\[ \{ S_{2N-1}^2, \Omega _+ \}_1  + R_{2N-1}^2 =0 ~.\]
We see that $ S_{2N-1}^2$ is known at this step, while
$ S_{2N-1}^0$ is modulo known terms a solution of
\[ \{\Omega _+, S_{2N-1}^0    \}_1= ax^{2N-1}~.\]
This equation gives, always mod known terms:
\[  S_{2N-1}^0  =a\Sigma _{2N-1} ~, \]
with $\Sigma _{2N-1}$ given by Definition \ref{def:sigma}.
\item 
{\bf In degree $2N$:}
\begin{eqnarray*}\label{2N}
bx^{2N}+ \frac{i}{\hbar}\left( [S_3,ax^{2N-1}]^\star+
[S_{2N},\Omega _+ ]^\star + [S_{2N-1},a_3 x^3]^\star \right)+\\
+\ha \left( \frac{i}{\hbar} \right)^2\left(
[S_{2N-1},[S_3,\Omega _+]^\star]^\star+[S_{3},[S_{2N-1},\Omega
_+]^\star]^\star
\right) +R_{2N}=B_{2N} +O(2N+1)
~.\nonumber \end{eqnarray*}
The previous equation  gives one equation in $\hbar^0 $
and one in $\hbar ^2$:
\begin{itemize}
%-----------------------------------------------------------
%degree 2N, hbar^0
\item {\bf degree $2N$, hbar$^0$}
\begin{eqnarray*}
bx^{2N}+\{ S_3, ax^{2N-1} \}_1 +\{ S_{2N}^0,\Omega _+ \}_1
+\{ S_{2N-1}^0,a_3x^3 \}_1 +\\
+\ha \{S_{2N-1}^0 ,\{S_3 ,\Omega _+ \}_1 \}_1
+\ha \{S_3 ,\{S_{2N-1}^0 ,\Omega _+ \}_1 \}_1
+R_{2N}^{0} =b_{2N}^0\Omega _+^N 
\nonumber \end{eqnarray*}
This can be simplified as:
\begin{equation}
\label{hbar0}  \{ \Omega _+ ,S_{2N}^0\}_1 =-b_{2N}^0 \Omega _+^{N}
+bx^{2N}+ \frac{a}{2}\{ S_3 ,x^{2N-1} \}_1 +\frac{a_3}{2}
\{ S_{2N-1}^0 , x^3 \}_1 +R_{2N}^{0}-\ha \{ S_3, R_{2N-1}^0\}_1 ~
\end{equation}
This gives 
$b_{2N}^0=\beta _N b + \gamma _N a_3 a $
modulo known terms. 
Moreover, Lemma \ref{lemm:x2N} implies  $\beta_N\ne 0 $. 
%-----------------------------------------------------------
%degree 2N, hbar^2
\item  {\bf degree $2N$, hbar$^2$}
\begin{eqnarray*}
-\frac{1}{24}\left( \{ S_3, ax^{2N-1}\}_3 + \{ S_{2N-1}^0,
  a_3x^{3}\}_3 \right)  +
\{ S_{2N}^2, \Omega _+ \}_1+  \\
+ \ha \left( \{ S_{2N-1}^2, \{ S_3 ,\Omega _+ \}_1\}_1+ 
\{ S_3, \{ S_{2N-1}^2, \Omega _+  \}_1\}_1
\right)- \nonumber \\
-\frac{1}{48}\left( \{ S_{2N-1}^0, \{ S_3 ,\Omega _+ \}_1\}_3+ 
\{ S_3, \{ S_{2N-1}^0, \Omega _+  \}_1\}_3
\right)
 + R_{2N}^{2}=b_{2N}^2\Omega _+^{N-2} 
\nonumber \end{eqnarray*}
which can be simplified as:
\begin{equation} \label{hbar2}\{ \Omega _+ ,  S_{2N}^2  \}_1=
-b_{2N}^2 \Omega_+^{N-2}
 -\frac{1}{48} \left(
a\{ S_3,x^{2N-1}\}_3 +a_3  \{ S_{2N-1}^0  ,x^3 \}_3 \right)
+R_{2N}^2 ~, 
\end{equation} modulo known terms.
This gives, using Lemma \ref{lemm:homol}, 
$ b_{2N}^2 =\gd _N a a_3 $ modulo known  terms.
\end{itemize}
\end{itemize}

From Equation (\ref{hbar2}) and the expressions for
$S_3$ (Equation (\ref{equ:S3})) and $\Sigma _{2N-1}$
(Lemma \ref{lemm:Sigma}),  we get:
\begin{equation} \label{deltaN}
\{ \Omega _+ ,  S_{2N}^2  \}_1=
aa_3 \frac{(N-1)(2N^2-4N +3)}{3}x^{2N-4}
-b_{2N}^2 \Omega_+^{N-2} {\rm ~mod~known~terms}
\end{equation}
Because $x^{2N-4} $ is not a Poisson bracket with
$\Omega _+$ by Lemma \ref{lemm:x2N},  we get
$\gd_N \ne 0$.

From the fact that $\gb_N$ and $\gd_N$ do not vanish,
this concludes the induction $N-1 \ra N$.

\section{Getting the QBNF from the density of states
in case of a local extremum of the potential}
\label{sec:density}

\subsection{$\hbar$-dependent distributions}
Let $T_\hbar $ be  an $\hbar$-dependent Schwartz distribution
on an open interval 
$J $.
\begin{defi}
 The family  $T_\hbar $ is
\begin{itemize}
\item  {\rm  regular} at the point $E_0 \in J$
if there exists 
a  sequence of functions
$T_j $ which are smooth in some
neighbourhood $K$  of $E_0$ with $j=j_0,~j_0+1,\cdots $
($j_0\in\Z$),  
so that, for any $f\in C_o^\infty (K)$,
we have the asymptotic expansion 
$T_\hbar (f) \equiv \sum _{j=j_0}^{+\infty} \hbar ^j \int _J T_j(x)f(x)
dx $.
\item  {\rm right regular}
(resp.  {\rm left regular})
  at the point $E_0 \in J$
if there exists  $E_1 > E_0$ (resp. $E_1<E_0$)
and a  sequence of functions
$T_j $ which are smooth in some
neighbourhood  of $E_0$ with $j=j_0,~j_0+1,\cdots $
($j_0\in\Z$),  
so that, for any $f\in C_o^\infty (]E_0, E_1[)$,
we have the asymptotic expansion 
$T_\hbar (f) \equiv \sum _{j=j_0}^{+\infty} \hbar ^j \int _J T_j(x)f(x)
dx $.
\end{itemize}
\end{defi}
We will use the following notations:
\begin{defi}
If $T_\hbar $ is a family of distributions on $J$
and $E_0 \in J$, $T_\hbar ^+$ (resp.  $T_\hbar ^-$),
the {\rm right (resp left) singular part of } $T_\hbar $ is
the equivalence class  of  $T_\hbar $ modulo families of distributions
which are right-(resp. left-)regular at the point $E_0$.
\end{defi}

\subsection{Density of states}
Consider a smooth potential $V:I\ra \R$ where $I$ is an open
interval with $ 0\in I$ 
and $\liminf  _{x\ra \pa I}V(x)=E_\infty > -\infty  $
and let $\hat{H}$ be the Schr\"odinger operator with potential $V$.
\begin{defi}
The {\rm density of states } is the $\hbar-$dependent 
Schwartz distribution $T_\hbar $  on
$]-\infty ,E_\infty [$ defined by 
\[ D_\hbar (f):={\rm Trace}f( \hat{H})
 ~.\]
\end{defi}

\begin{lemm}\label{lemm:regular}
If $J$ is an open subset of $  ]-\infty ,E_\infty [$
which  contains no critical values
of $V$, the density of states is regular at every point of $J$.
\end{lemm}
\begin{demo}
Let us denote by $H=\ha \xi^2 +V(x)$ the symbol of the
Schr\"odinger operator  $\hat{H}$.
The operator $f(\hat{H})$ is a pseudo-differential operator
whose symbol ${f}^\star ({H})$ is given (see \cite{CV2}) by:
 \[ {f}^\star({H})=f(H)+ \sum _{j\geq 1,~l\geq 1}
\hbar^{2j}P_{j,l}(x,\xi)f^{(l)}(H) ~,\]
where the $P_{j,l}$'s are smooth functions locally computable from
the symbol $H$.
It is now enough to check that
$f\ra (2\pi \hbar)^{-1}\int\!\int  P_{j,l}(x,\xi)f^{(l)}(H)dx d\xi $
is regular at each point of $J$ using the fact that $H$ has no
critical
value in $J$. 
\end{demo}

\subsection{Singularity of the density of states
near a local maximum of the potential}

Let us assume that 
$V(0)=E_0< E_\infty ,~ V'(0)=0$ and $V''(0)<0$.
 Assume also that $0$ is the 
unique critical point of $V$ whose  critical value is  $E_0$.

We have the:
\begin{theo} \label{theo:singu}
If the QBNF of $\hat{H}$
is \[ B \equiv
\Omega _- +\sum _{2j+k\geq 2}b_{j,k}\hbar^{2j}\Omega _-^k ~,\]
the density of states is right and left singular at the point $E_0$
 and
one can recover the full QBNF (the coefficients $b_{j,k}$)
from the right (resp. left) singular part $ D_\hbar^+ $
(resp.  $ D_\hbar^- $)of the density of states   $ D_\hbar $  at $E_0$.
\end{theo}
In what follows, it is more convenient to use
$\Omega_- =x\xi$.
\subsubsection{The singularity of the density of states and the QBNF}

\begin{lemm} If $B$ is the QBNF of
$\hat{H}$, the singular part of the density of states
is the same as that of the family of distributions
\[ G_\hbar : f \ra \frac{1}{2\pi \hbar}
\int \! \int _D f^\star (B) dx d\xi ~,\]
where $f^\star (B)$ is the Weyl symbol of 
$f(\hat{B})$ and $D $ is the square
$\max (|x| ,| \xi |) \leq 1 $.
\end{lemm}

\begin{demo}
Let $\Pi={\rm Op}_{\rm Weyl}(\go ) $ be a compactly supported
 \OPD ~whose Weyl symbol $\go $ is 
$\equiv 1 $ near $(0,0)$.
We have
\[ D_\hbar (f)=(2\pi \hbar )^{-1}\left(  \int \! \int \go \star f^\star
  (H)dx d\xi 
+ \int \! \int (1-\go ) \star f^\star (H)  dx d\xi \right)~.\]
Using (the proof of) Lemma \ref{lemm:regular},
the second term is a regular distribution.
The first term can be transformed using the QBNF:
there exists an FIO $U$, microlocally unitary,
which transforms $\hat{H}$ into its QBNF and hence for every function
$f$, we have:
\[ U^\star f(\hat{B})U = f(\hat{H}) \] 
microlocally near the origin.
In this way, we get
\[ {\rm Trace}(\Pi \circ f(\hat{H}))\equiv
 {\rm Trace}(\Pi U^\star f(\hat{B}) U )~.\]
Introducing 
$\Pi_1:=U \Pi U^\star $
(a \OPD ~whose Weyl symbol is $\equiv 1$
near the origin)  and  using the commutativity  of the trace,
we have:
 \[ {\rm Trace}(\Pi \circ f(\hat{H}))\equiv
 {\rm Trace}(\Pi_1 \circ f(\hat{B})))~.\]
It remains to check that, if $ \Pi_1={\rm Op}_{\rm Weyl}(\omega_1)$,
$f \ra \int \! \int _{(I\times \R) \setminus D}
\omega _1 \star f^\star (B) dx d\xi $
is regular.
\end{demo}

\subsubsection{Computing some singularities}

\begin{lemm} \label{lemm:f(B)}
Let us consider the family of distributions
\[ K_\hbar (f)=\int \! \int _D f\left(\sum _{j=0}^\infty \hbar
^{2j}b_j(x\xi)\right) dx d\xi \]
 on $]E_0, E_1[$ (we consider only the case  $E_1>E_0$, the other
case is similar),
where the $b_j$'s are smooth on $[E_0, E_1 ]$ and $b_0 (u) \equiv E_0 +
\sum_{j=1}^\infty
 \beta _j u^j$ with  $\beta _1 > 0 $.
Then  $K_\hbar (f) $ admits an asymptotic expansion in powers
of $\hbar $:
\[  K_\hbar (f) \equiv \sum _{j=0}^\infty K_j(f) \hbar ^{2j} \]
and the  right singularities    of $K_0,\cdots , K_N$
at the point $E_0$  determine
the Taylor expansions of $b_0, \cdots ,b_N$ at the origin. 
\end{lemm}

\begin{demo}
Let us Taylor expand the integrand as:
\[ \begin{array}{rl} f\left(\sum _{j=0}^\infty \hbar
^{2j}b_j(x\xi)\right)&\equiv f(b_0(x\xi))+f'(b_0(x\xi))
\left( \sum _{j=1}^\infty  \hbar ^{2j}b_j(x\xi) \right)\\
  +& \sum _{k=2}^\infty  \frac{1}{k!}
f^{(k)}(b_0(x\xi))
\left( \sum _{j=1}^\infty  \hbar ^{2j}b_j(x\xi) \right)^k~.
\end{array}~,\]
\[ \equiv 
 f(b_0(x\xi)) +\sum _{j=1}^\infty  \hbar ^{2j}\left(
f'(b_0(x\xi))b_j(x\xi) + \sum _l f^{(l)}(b_0 (x\xi)) R_{j,l}(x\xi ) \right)~,\]
where the functions $R_{j,l}$ depend only on 
$b_1,\cdots ,b_{j-1}$.

We have to prove the following 2 facts:
\begin{enumerate}
\item The right singularity of
\[ \int \! \int _D  f(b_0(x\xi))dxd\xi \]
determines the Taylor expansion of $b_0$ at the origin.
\item  The right singularity of
\[ \int \! \int _D  f'(b_0(x\xi))b_j(x\xi)dxd\xi \]
determines the Taylor expansion of $b_j$ at the origin,
assuming the Taylor expansion of $b_0$ is  known.
\end{enumerate}
Both are easy consequences of the following elementary calculus result:
\[ \int \! \int _D  f'(b_0(x\xi))b_j(x\xi)dxd\xi
\equiv \int _{E_0}^{E_1} f'(t) b_j (c_0(t))c'_0(t) |\log (t-E_0)|
dt ~ \]
(modulo smooth distributions) where $c_0$ is the inverse function
of $b_0$.

\end{demo}

\subsubsection{End of the proof of Theorem \ref{theo:singu}}

We have 
\[ f^\star (B)(z_0)=\sum _{j=0}^\infty \frac{1}{2j!}
\left( f^{(2j)}(B(z_0)) (B-B(z_0))^{\star 2j } \right) (z_0)~. \]
It is enough to check the:
\begin{lemm}
If \[ f^\star (B)\equiv f(B)+ \sum _{j=1}^\infty
\hbar ^{2j} \sum_l  f^{(2l)}(B) R_{j,l} ~,\]
the $R_{j,l}$'s depend only on 
$b_0,\cdots , b_{j-1}$.
\end{lemm}
\begin{demo}
The $\star -$powers of $B-B(z_0)$ evaluated at $z_0$ start 
with terms in $\hbar ^2$ and the $b_l$'s, for $l\geq j$ have already
an $\hbar ^{2j}$ in front of them!
\end{demo}

So everything works as if $f^{\star}(B)=f(B) $ and we are reduced to
Lemma
\ref{lemm:f(B)}.

\subsection{The case of a local minimum} \label{sec:minimum}

The same strategy applies, but now the density of states is right AND
left regular, with a jump singularity at the point $E_0$.

We get:
\begin{theo} \label{theo:min_loc}
If the QBNF of $\hat{H}$
is \[ B \equiv
\Omega _+ +\sum _{2j+k\geq 2}b_{j,k}\hbar^{2j}\Omega _+^k ~,\]
the density of states is  singular at the point $E_0$
 and
one can recover the full QBNF (the coefficients $b_{j,k}$)
from the  singular part of the density of states   $ D_\hbar $  at $E_0$.
\end{theo}

The proof is very similar to the case of a local 
maximum. We have now a ``Heaviside singularity'', meaning that
the density of states is right AND
left regular, but  the functions 
$T_j$ defined by 
\[ D_\hbar (f)= \sum _{j=-1}^\infty \int fT_j \hbar ^j ~\]
 and their
derivatives
have jumps 
at the point $E_0$.
We have only to look at the singularities of 
$T:f\ra \int f(\Omega _+) dx d\xi $.
We have $T(f)=2\pi \int _0^{+\infty} f(u) du $, so
$T=2\pi Y $ where $Y$ is the Heaviside function.

\section{Open problems}

\begin{itemize}
\item {\bf Is the result still true if $a_3=0$?}
This  is the case modulo some global assumption
on $V$  (see  \cite{CV3}). In fact in \cite{CV3}, it is 
shown that, modulo some genericity assumptions,  
the potential itself is determined from its semi-classical spectrum.
% It is already valid for the first 
%coefficients as a consequence of the calculations in Appendix A.
\item {\bf Is the result still valid in any dimension?}
  We think that the answer is no, at least it does not work with the
  same
arguments; let us assume that the quadratic part of the Hamiltonian is
 $H_2=\go_1 \Omega _1 +\go _2 \Omega _2 $
with $\Omega _1 $ (resp. $\Omega _2$) harmonic oscillators in $x_1$
(resp. $x_2$).
\begin{itemize}
\item Non resonant case: $\go _1 $ and $\go _2$ are independent over 
$\Z$.
 In degree $4$,  the QBNF has $4$ unknown 
coefficients, an homogeneous polynomial of degree
$2$ in $(\Omega _1,\Omega _2) $ and the coefficient of $\hbar ^2$.
On the other hand,
 $V_3(x_1,x_2)+V_4(x_1,x_2)$ has $9(>4)$ coefficients.
However, it is possible that higher terms in the QBNF
give other information's...
\item Resonant case: $\go_1=\go _2$.
In degree $4$, the classical BNF has already $9$
coefficients (it is a polynomial of degree
$4$ on $\R^4$ invariant by the
circle action generated by the flow of $\Omega _1 + \Omega _2$),
 this seems promising. However, we have to
take into account an $O(2)$ action by isometries 
in $\R^2$: on one hand, we can only expect to determine
the potential up to this action; on the other hand,
the QBNF is determined only up to action
by $SU(2)$. 
\end{itemize}
\end{itemize}
\section{Homogeneity properties of the QBNF}

We have the following:
\begin{theo}
The $b_{j,k}$'s (coefficients of
$\hbar^{2j}\Omega ^k $ in the QBNF) satisfy the following
 homogeneity properties:
\[ b_{j,k}(ta_3,t^2a_4,\cdots ,t^n a_{n+2},\cdots)=
t^{2(2j+k)-2}b_{j,k}(a_3, a_4,\cdots )~.\]
\end{theo}
\begin{demo}
Let us consider 
\[ \hat{H}_t=\ha \left( -\hbar^2 \frac{d^2}{dx^2}+x^2 \right)+
ta_3x^3 +\cdots + t^{n-2} a_n x^n +\cdots ~,\]
and make the change of variable
$tx=y,~\hbar_1=t^2 \hbar $.
We get a new operator
\[t^{-2}\left[ \ha \left( -\hbar_1^2 \frac{d^2}{dy^2}+y^2 \right)+
a_3y^3 +\cdots +  a_n y^n +\cdots \right]~.\]
The spectrum   of the second one is then $t^{-2}$times that of the
first one.
This implies the property.

\end{demo}
%\section*{Appendix A: calculations of the QBNF by Fr\'ed\'eric Faure}

\bibliographystyle{plain}

\end{document}